\newcommand{\green}{f_{_{\rm G}}}
\documentclass{kluwer}    % Specifies the document style.
\usepackage{graphicx}
\newdisplay{guess}{Conjecture}
\begin{document}
\begin{article}
\begin{opening}         

\title{Noise storm continua: power estimates for electron acceleration}
\author{Prasad \surname{Subramanian}\thanks{e-mail: psubrama@iucaa.ernet.in}}  

\institute{Inter-University Centre for Astronomy and Astrophysics,
P.O. Bag 4, Ganeshkhind, Pune - 411007, India.}
\author{Peter A. \surname{Becker}\thanks{e-mail: pbecker@gmu.edu}
\thanks{also Dept. of Physics and Astronomy,
George Mason University, Fairfax, VA 22030, USA}}
\institute{Center for Earth Observing and Space Research, \break School of
Computational Sciences, \break George Mason University, Fairfax,
VA 22030, USA}
\runningauthor{Subramanian \& Becker}
\runningtitle{Noise storm continua}

%\date{}

\begin{abstract}
We use a generic stochastic acceleration formalism to examine the power
$L_{\rm in}$ (${\rm erg\,s^{-1}}$) input to nonthermal electrons that
cause noise storm continuum emission. The analytical approach includes
the derivation of the Green's function for a general second-order Fermi
process, and its application to obtain the particular solution for the
nonthermal electron distribution resulting from the acceleration of a
Maxwellian source in the corona. We compare $L_{\rm in}$ with the power
$L_{\rm out}$ observed in noise storm radiation. Using typical values
for the various parameters, we find that $L_{\rm in} \sim 10^{23-26}$
${\rm erg\,s^{-1}}$, yielding an efficiency estimate $\eta \equiv L_{\rm
out}/L_{\rm in}$ in the range $10^{-10} \lsim \eta \lsim 10^{-6}$ for
this nonthermal acceleration/radiation process. These results reflect
the efficiency of the overall process, starting from electron
acceleration and culminating in the observed noise storm emission.
\end{abstract}
\keywords{}

\end{opening}

\section{Introduction} 
Solar noise storms are a very well studied phenomenon. They occur
mostly at meter wavelengths, and comprise a long-lasting (1 hr --
several days) broadband ($\delta f/f \sim 100\,\%$) component together
with narrowband ($\delta f/f \sim {\rm few}\,\%$) spiky bursts that last
from 0.1 -- 1 s. Elgaroy (1997) gives a thorough observational review of
noise storms. Malik \& Mercier (1996) present a comprehensive study of
noise storms observed with the Nancay Radioheliograph (NRH). Klein
(1998) presents a recent review of the role of suprathermal electrons
in the solar corona that includes a broad discussion of the noise storm
phenomenon.

In this study, we confine our attention primarily to type I noise storm
{\em continua}, rather than the sporadic type I bursts, because we are
interested in examining the basic energetics of the electron
acceleration processes responsible for producing the quasi-continuous
radio emission.

Most theories of type I phenomena invoke nonthermal electrons as a
crucial ingredient in producing the observed radiation. It is recognized
that the nonthermal electrons that are involved in generating the noise
storm continua are probably accelerated in closed coronal loops above
active regions. Noise storm continua are sometimes accompanied by
coincident (thermal) soft X-ray brightenings (e.g., Raulin \& Klein
1994; Krucker et al. 1995), which are also signatures of coronal
magnetic field evolution. When this occurs, it is clear that the
electrons in the tail of the thermal distribution that produce the
accompanying soft X-radiation cannot also produce the observed radio
noise storm emission for more than a few minutes. This is inconsistent
with the fact that noise storms are observed to persist for several
hours to days (e.g., Raulin \& Klein 1994; Malik \& Mercier 1996 also
present similar arguments). Since the X-ray emitting regions are
typically situated at least one scale height below the layer in the
corona where the noise storms originate, the two associated electron
distributions cannot be co-spatial. Crosby et al. (1996) find that
deka-keV X-ray emission is often observed towards the beginning of noise
storms. The electrons producing this X-ray emission are energetic enough
to power the noise storm simultaneously, and they could be transported
to the noise storm emitting region either by turbulent diffusion or by
direct transport along connecting magnetic field lines. However, the
duration of the noise storm is much longer than the X-ray emission, and
consequently continual electron acceleration is required. The
acceleration is probably triggered by the same processes that give rise
to the X-ray brightenings accompanying the onset of the noise storms.

Clearly an underlying acceleration mechanism is required in order to
produce the nonthermal electron distribution implied by the noise storm
emission. However, very little attention has been focused on this
problem in the previous literature. The majority of the current theories
simply assume that nonthermal electrons are present, and focus most of
their attention on examining the wave-wave interaction processes through
which observable radio emission is ultimately produced. It is fairly
well established that noise storms are intimately connected with the
temporal variation of the magnetic fields via the process of coronal
evolution (e.g., Brueckner 1983; Stewart et al. 1986; Raulin \& Klein
1994; Willson et al. 1997; Bentley et al. 2000). However, the precise
physical processes that link the magnetic field evolution to the
presence of nonthermal electrons is unclear. Spicer et al. (1981)
adopted a specific driver model in order to examine the basic energetics
of the process leading to type I bursts. Benz \& Wentzel (1981)
developed a similar treatment that is also valid for type I bursts. The
basic driver in both of these pictures is the process of coronal
evolution, which causes magnetic fields to emerge into the corona and
drive microinstabilities that spawn low-frequency turbulence. This
turbulence in turn resonates with the electrons and stochastically
accelerates them to form a nonthermal tail.

Our goal here is to develop a model-independent approach to the
estimation of the noise storm energetics that avoids focusing on a
specific mechanism for accelerating the electrons. Instead, we prescribe
a generic second-order (stochastic) Fermi mechanism. Electrons from the
tail of the thermal distribution are injected into the acceleration
process, forming a nonthermal distribution. Only electrons above a
critical energy in the tail of the thermal distribution are subjected to
net acceleration; the rest remain thermal due to collisions. The basic
parameters of the acceleration mechanism are constrained as follows.
First we borrow from the literature estimates of the nonthermal electron
fraction needed to produce the observed noise storm continua. By
combining this estimate with approximate expressions for the dominant
loss timescales influencing the electrons, we derive an estimate for the
power that drives the electron acceleration process. This yields an
approximate determination of the efficiency of the process starting from
nonthermal electron acceleration and culminating in the observed noise
storm emission. The efficiency so obtained is a relatively well-defined
quantity that provides a general, model-independent constraint on the
acceleration/radiation mechanisms, which in turn serves as a useful
guide in the subsequent development of more detailed models.
Furthermore, we expect that a good understanding of the efficiency of
electron acceleration in the context of the solar corona will also
provide useful insights into similar phenomena occurring in other
astrophysical environments.

\section{Electron acceleration}

\subsection{Why are nonthermal electrons important?}

There is considerable observational evidence for the presence of
nonthermal electrons in nonflaring regions of the solar corona (e.g.,
Klein 1998). The high brightness temperatures and significant positive
spectral slopes in multi-frequency observations of noise storms strongly
suggest an underlying nonthermal electron population (e.g., Thejappa \&
Kundu 1991; Sundaram \& Subramanian 2004). An anisotropy in velocity or
physical space causes these nonthermal electrons to spontaneously emit
Langmuir/upper hybrid waves (e.g., Robinson 1978; Wentzel 1985), which
can coalesce with a suitable low-frequency wave population to produce
the observed electromagnetic emission.

Melrose (1980) argues that low-frequency turbulence (including
ion-acoustic, lower-hybrid, or a variety of other waves) will be
generated as a natural product of coronal heating. Thejappa (1991)
considers lower-hybrid waves excited by protons (as in Wentzel 1986) or
by a series of weak shocks (as in Spicer et al. 1981) as candidates for
the low-frequency wave population. Benz \& Wentzel (1981) investigate
the possibility that the turbulence is composed of ion-acoustic waves
excited by evolving magnetic fields in the corona.

While most of the theories for noise storm continua do not specify the
source of the nonthermal electrons powering the observed emission, Benz
\& Wentzel (1981) propose that these electrons are leftover particles
from the population that produced the previous type I bursts. However,
Krucker et al. (1995) note that the type I continuum and bursts are
spatially separated. Furthermore, Malik \& Mercier (1996) pointed out
that solar noise storms are often not bursty in the beginning, and the
continuum exists alone. These observational results contradict Benz \&
Wentzel's (1981) picture for the production of the nonthermal electrons.
On the other hand, the model proposed by Spicer et al. (1981) for type I
bursts considers the electrons to be accelerated via the modified
two-stream instability, which in turn is spawned by random weak shocks
caused by the emerging magnetic flux. However, this suggestion is
contradicted by the work of Krucker et al. (1995).

Although there is currently no theoretical consensus regarding the
fundamental mechanism powering type I phenomena, there is no question
that nonthermal electrons are responsible for the observed emission.
This situation leads us to suggest that a careful examination of the
energy budget and the associated constraints on the efficiency of the
acceleration process may provide the best route towards enhanced
physical understanding.

\subsection{Energy loss mechanisms}

It is expected that the electron momentum distribution, $f$, will have
essentially a two-part structure, comprising a thermal component along
with a nonthermal, high-energy (but still nonrelativistic) electron
``tail'' that is responsible for producing the noise storm emission. In
view of the considerable uncertainty surrounding the precise physical
mechanism that results in the formation of the nonthermal portion of the
electron distribution, we shall work in terms of a generic, stochastic
(second-order) Fermi process. The total electron number density, $n_e$,
including both thermal and nonthermal particles, is related to the momentum
distribution $f$ via
\begin{equation}
n_e \, ({\rm cm^{-3}}) \, = \, \int_0^\infty p^2 \, f \, dp \ ,
\label{eq1}
\end{equation}
where $p$ is the electron momentum. The associated total electron energy
density is given by
\begin{equation}
U_e \, ({\rm erg\,cm^{-3}}) \, = \, \int_0^\infty \epsilon \, p^2
\, f \, dp = \frac{1}{2 m_e} \, \int_0^\infty \, p^4 \, f \, dp \ ,
\label{eq2}
\end{equation}
where $m_e$ is the electron mass and $\epsilon = p^2/(2 m_e)$ is
the electron kinetic energy. In the present application, we are mainly
interested in the nonthermal electrons, which are picked up from the
thermal population and subsequently accelerated to high energies.

At marginal stability, the loss timescale due to the emission of
Langmuir waves by electrons is similar to the Coulomb loss timescale
(Melrose 1980). The mean Coulomb energy loss rate for nonthermal
electrons with energy $\epsilon$ colliding with ambient electrons and
protons in a fully-ionized hydrogen plasma is given in cgs units by
(e.g., Brown 1972b)
\begin{equation}
\bigg\langle{d\epsilon \over dt}\bigg\rangle \bigg|_{\rm loss}
= {1.57 \times 10^{-23} \, \Lambda \, n_e \over \epsilon^{1/2}}
\left(1 + {m_e \over m_p}\right)
\ ,
\label{eq3}
\end{equation}
where $\Lambda$ is the Coulomb logarithm and $m_p$ is the proton mass.
The first and second terms in parentheses on the right-hand side of
equation~(\ref{eq3}) refer to electron-electron and electron-proton
collisions, respectively. The factor of $m_e/m_p$ clearly indicates that
the cooling rate due to electron-proton bremsstrahlung is negligible
compared with the effect of electron-electron collisions. Ignoring the
factor of $m_e / m_p$ in equation~(\ref{eq3}), we find that the
characteristic timescale for Coulomb losses is given by
\begin{equation}
t_{\rm loss} \equiv \epsilon \ \bigg\langle{d\epsilon \over dt}
\bigg\rangle^{-1} \bigg|_{\rm loss}
= {6.37 \times 10^{22} \, \epsilon^{3/2} \over \Lambda \, n_e}
\label{eq5}
\end{equation}
in cgs units.

\subsection{Stochastic acceleration of electrons}

For low electron energies, $\epsilon \lsim kT$, collisions between
electrons are expected to efficiently maintain a Maxwellian
distribution. However, we shall demonstrate below that for electron
energies exceeding the critical energy $\epsilon_c$, stochastic
acceleration dominates over collisional losses on average (see
eq.~[\ref{eq10}]). In this situation, the high-energy tail of the
electron distribution function is governed by a rather simple transport
equation that describes the diffusion of electrons in momentum space due
to collisions with magnetic scattering centers. The time evolution of
the Green's function for this process, $\green$, is described by (e.g.,
Becker 1992; Schlickeiser 2002)
\begin{equation}
{\partial \green \over \partial t} =
\frac{1}{p^2}\,\frac{\partial}{\partial p}\left( p^2 \, {\cal D}\,
\frac{\partial \green}{\partial p}\right) + {\dot N_0 \, \delta(p-p_0)
\over p_0^2} - {\green \over \tau} \ ,
\label{eq6}
\end{equation}
where ${\cal D}$ is the (as yet unspecified) diffusion coefficient in
momentum space and $\tau$ is the mean residence time for electrons in
the acceleration region. The source term in equation~(\ref{eq6})
corresponds to the injection into the acceleration region of $\dot N_0$
particles per unit volume per unit time, each with momentum $p_0$.
Although we do not explicitly include losses due to the emission of
Langmuir/upper hybrid waves by the accelerated electrons, it is expected
that these waves will be generated as a natural consequence of the
spatial anisotropy of the electron distribution (e.g., Thejappa 1991).
We will demonstrate below that our neglect of energy losses due to
Coulomb collisions and the emission of Langmuir/upper hybrid waves is
reasonable for the nonthermal electrons treated by equation~(\ref{eq6}).
Note that in writing equation~(\ref{eq6}), we have ignored spatial
transport so as to avoid unnecessary mathematical complexity.

Although the specific form for ${\cal D}$ as a function of $p$ depends
on the spectrum of the turbulent waves that accelerates the electrons
(Smith 1977), it is possible to make some fairly broad generalizations
that help to simplify the analysis. In particular, we point out that a
number of authors have independently suggested that ${\cal D} \propto
p^2$. Examples include the treatment of particle acceleration by
large-scale compressible magnetohydrodynamical (MHD) turbulence (Ptuskin
1988; Chandran \& Maron 2003); analysis of the acceleration of electrons
by cascading fast-mode waves in flares (Miller, LaRosa \& Moore 1996);
and the energization of electrons due to lower hybrid turbulence (Luo et
al. 2003). Hence we shall write
\begin{equation}
{\cal D} = D_0 \, p^2 \, ,
\label{eq7}
\end{equation}
where $D_0$ is a constant with the units of inverse time.

We can obtain an expression for the mean rate of change of the electron
momentum $p$ due to stochastic acceleration by focusing on the instantaneous
evolution of a localized $\delta$-function distribution in momentum space.
Following the procedure described by Subramanian, Becker, \& Kazanas (1999),
we find based on equation~(\ref{eq6}) that the mean rate of change of the
particle momentum due to stochastic acceleration is given by
\begin{equation}
\bigg\langle{dp \over dt}\bigg\rangle \bigg|_{\rm accel}
= {1 \over p^2} \, {d \over dp} \left(p^2 \, {\cal D}\right)
= 4 \, D_0 \, p \ .
\label{eq8}
\end{equation}
Since $\epsilon = p^2/(2 \, m_e)$, we conclude that the corresponding mean
rate of change of the electron energy is
\begin{equation}
\bigg\langle{d\epsilon \over dt}\bigg\rangle \bigg|_{\rm accel}
= {p \over m_e} \, \bigg\langle{dp \over dt}\bigg\rangle
\bigg|_{\rm accel}
= 8 \, D_0 \, \epsilon \ .
\label{eq8b}
\end{equation}
The associated timescale for stochastic acceleration is therefore
\begin{equation}
t_{\rm accel} \equiv \epsilon \ \bigg\langle{d\epsilon \over dt}
\bigg\rangle^{-1} \bigg|_{\rm accel}
= {1 \over 8 \, D_0} \ .
\label{eq9}
\end{equation}
On average, acceleration dominates over losses for a particle with
energy $\epsilon$ if $t_{\rm accel} < t_{\rm loss}$. Comparison of
equations~(\ref{eq5}) and (\ref{eq9}) establishes that acceleration is
dominant if $\epsilon > \epsilon_c$, where the critical energy
$\epsilon_c$ is given in cgs units by
\begin{equation}
\epsilon_c \equiv 1.57 \times 10^{-16} \left(\Lambda \, n_e
\over D_0\right)^{2/3}
\ .
\label{eq10}
\end{equation}
It should be noted that due to the stochastic nature of the acceleration
process, some particles with energy $\epsilon > \epsilon_c$ will lose
energy, although on average such particles will gain energy. Our
assumption that the mean acceleration rate dominates over losses for the
nonthermal particles is self-consistent provided the nonthermal
particles all have $\epsilon > \epsilon_c$. The corresponding result for
the critical momentum is
\begin{equation}
p_c \equiv (2 \, m_e \, \epsilon_c)^{1/2}
= 1.64 \times 10^{-21} \left(n_e \over D_0\right)^{1/3}
\ ,
\label{eq11}
\end{equation}
where we have set the Coulomb logarithm $\Lambda = 29.1$ (see Brown
1972a). Particles with $p > p_c$ experience net acceleration on average.
Note that the electron source term appearing in equation~(\ref{eq6})
must have $p_0 > p_c$ in order to validate our neglect of collisional
losses in that equation. When this condition is satisfied, the injected
electrons are accelerated to form a nonthermal distribution. While we
have considered only Coulomb losses in this calculation, we remind the
reader that the loss timescale due to the emission of Langmuir waves at
marginal stability is similar to the Coulomb loss timescale (see
\S~2.2).

\subsection{Solution for the Green's function}

In a steady state, it is straightforward to show based on
equation~(\ref{eq6}) that when ${\cal D} = D_0 \, p^2$ as assumed here,
the solution for the Green's function is given by (Subramanian, Becker, \&
Kazanas 1999)
\begin{equation}
\green(p,p_0) = A_0 \, \cases{
(p/p_0)^{\alpha_1} \ , & $p \le p_0$ \ , \cr
\phantom{space} \cr
(p/p_0)^{\alpha_2} \ , & $p \ge p_0$ \ , \cr 
}
\label{eq12}
\end{equation}
where the exponents $\alpha_1$ and $\alpha_2$ are related to $D_0$ and
$\tau$ via
\begin{equation}
\alpha_1 \equiv - {3 \over 2} + \left({9 \over 4} + {1 \over D_0 \, \tau}
\right)^{1/2} \ , \ \ \ \ \
\alpha_2 \equiv - {3 \over 2} - \left({9 \over 4} + {1 \over D_0 \, \tau}
\right)^{1/2} \ ,
\label{eq13}
\end{equation}
and the normalization parameter $A_0$ has the value
\begin{equation}
A_0 = {\dot N_0 \over 2 \, D_0 \, p_0^3} \left({9 \over 4} +
{1 \over D_0 \, \tau}\right)^{-1/2} \ .
\label{eq14}
\end{equation}
Since the second-order Fermi acceleration process is stochastic in
nature, the particles diffuse away from the injection momentum $p_0$,
and the $p \ge p_0$ and $p \le p_0$ branches of the Green's function
appearing in equation~(\ref{eq12}) describe the acceleration and
deceleration of the source particles, respectively. However, on average,
acceleration wins out over deceleration, as indicated by
equation~(\ref{eq8b}) which demonstrates that the mean acceleration rate
is positive. Substitution into the integral $\int_0^\infty p^2 \, \green
\, dp$ using equation~(\ref{eq12}) confirms that the number density of
the Green's function is equal to $\dot N_0 \tau$, as expected in this
steady-state situation. The values of $D_0$, $\tau$, and $\dot N_0$ will
be constrained later using observational data.

The Green's function $\green$ describes the response to the injection of
$\dot N_0$ electrons per unit volume per unit time with momentum $p_0$,
and therefore it is easy to show based on the linearity of the transport
equation~(\ref{eq6}) that the particular solution for a general
source term $j$ is obtained via the convolution (e.g., Becker 2003)
\begin{equation}
f(p) = \int_0^\infty {p_0^2 \, j(p_0) \over \dot N_0} \, \green
(p,p_0) \, dp_0 \ ,
\label{convolve}
\end{equation}
where the quantity $p_0^2 \, j(p_0) \, dp_0$ represents the number of
electrons injected per second per $\rm cm^3$ with momenta between $p_0$
and $p_0 + dp_0$. In the physical application of interest here, the
injected particles are supplied by the high-energy ($p_0 > p_c$) portion
of the Maxwellian distribution in the corona. Assuming that the
characteristic timescale for the thermal electrons to enter the
acceleration region is equal to the mean residence time $\tau$, it
follows that the source term is given by
\begin{equation}
j(p_0) = \cases{
{4 \pi n_e \, \tau^{-1} \over (2 \pi m_e k T)^{3/2}}
\ e^{-p_0^2 / 2 m_e k T} \ , & $p_0 \ge p_c$ \ , \cr
\phantom{space} \cr
0 \ , & $p_0 < p_c$ \ . \cr 
}
\label{source}
\end{equation}

\subsection{Particular solution for Maxwellian source}

In the case of a Maxwellian source, which is our focus here, we can
combine equations~(\ref{eq12}) through (\ref{source}) to compute the
particular solution for the nonthermal electron distribution. The lower
bound for the integration over $p_0$ in equation~(\ref{convolve}) is set
equal to $p_c$ since the source $j(p_0)$ vanishes for $p_0 < p_c$
according to equation~(\ref{source}). The result obtained for the
nonthermal electron distribution is therefore given by
\begin{equation}
f(p) =
{n_e \left\{
\xi^{\alpha_1/2} \, \Gamma\left(- \, {\alpha_1 \over 2} , \xi\right)
- \xi^{\alpha_2/2} \left[ \Gamma \left(- \, {\alpha_2 \over 2} , \xi
\right) - \Gamma \left(- \, {\alpha_2 \over 2} , \xi_c \right)
\right]\right\}
\over \sqrt{2 \pi} \, (m_e k T)^{3/2} \, (\alpha_1 - \alpha_2)
\, D_0 \, \tau}
\ ,
\label{particular}
\end{equation}
where we have introduced the dimensionless electron energy $\xi$ and
the dimensionless critical energy $\xi_c$, defined by
\begin{equation}
\xi \equiv {p^2 \over 2 \, m_e k \, T} \ , \ \ \ \ \ 
\xi_c \equiv {p_c^2 \over 2 \, m_e k \, T} \ .
\label{particular2}
\end{equation}
The number and energy densities associated with the particular solution
$f$ above the critical momentum $p_c$ can be computed using
\begin{equation}
n_* \, ({\rm cm^{-3}}) \, \equiv \, \int_{p_c}^\infty p^2 \, f(p)
\, dp \ ,
\label{eq15}
\end{equation}
\begin{equation}
U_* \, ({\rm erg\,cm^{-3}}) \, \equiv \, \int_{p_c}^\infty \epsilon
\, p^2 \, f(p) \, dp \ .
\label{eq16}
\end{equation}
By combining equations~(\ref{particular}), (\ref{eq15}), and (\ref{eq16}),
we find that the exact solutions for $n_*$ and $U_*$ are given by
\begin{eqnarray}
& {n_* \over n_e} =
{2 \, \xi_c^{(3+\alpha_1)/2} \, \Gamma\left(- {\alpha_1 \over 2} , \, \xi_c
\right) \over \sqrt{\pi} \, (3+\alpha_1) (\alpha_2-\alpha_1) \, D_0 \, \tau}
\nonumber \\
& \phantom{lotsofspaaace} + \, 2  \, e^{-\xi_c} \left(\xi_c \over \pi\right)^{1/2}
+ {\rm Erfc}\left(\xi_c^{1/2}\right)
\ ,
\label{eq15b}
\end{eqnarray}
\begin{eqnarray}
& {U_* \over n_e k \, T} =
{2 \, \xi_c^{(5+\alpha_1)/2} \, \Gamma\left(- {\alpha_1 \over 2} , \, \xi_c
\right) \over \sqrt{\pi} \, (5+\alpha_1) (\alpha_2-\alpha_1) \, D_0 \, \tau}
\nonumber \\
& \phantom{lotsofspaaace} + \, {2 \sqrt{\pi \xi_c} \, (3 + 2 \, \xi_c) \, e^{-\xi_c}
+ \, 3 \, \pi \, {\rm Erfc}\left(\xi_c^{1/2}\right) \over
2 \, \pi (1 - 10 \, D_0 \, \tau)} \ .
\label{eq16b}
\end{eqnarray}
Based on equations~(\ref{eq8b}) and (\ref{eq16}), we conclude that the
rate of change of the energy density of the nonthermal electrons
due to second-order Fermi acceleration is given by
\begin{equation}
\nonumber
{dU_* \over dt} \bigg|_{\rm accel}
= \int_{p_c}^\infty p^2 \, \bigg\langle{d\epsilon \over dt}
\bigg\rangle \bigg|_{\rm accel} \, f(p) \, dp
= 8 \, D_0 \, U_* \ ,
\label{eq17}
\end{equation}
with $U_*$ computed using equation~(\ref{eq16b}).

\section{Estimate of power input to electron acceleration process}

Assuming that the nonthermal electrons spontaneously emit Langmuir
waves, which then coalesce with lower hybrid waves to produce the
observable electromagnetic radiation, Thejappa (1991) gives estimates of
the fraction of nonthermal electrons, $n_* / n_e$, that are needed to
produce a given noise storm continuum brightness temperature $T_{b}$ at
a particular observing frequency. In this model, noise storm radiation
exhibits broadband continuum characteristics provided $n_{*} / n_e$ is
small enough so that collisional damping (due to ambient thermal
electrons) dominates over negative damping due to nonthermal electrons.
When $n_* / n_e$ exceeds a certain threshold, the negative damping due
to nonthermal electrons dominates, the brightness temperature increases
steeply and type I bursts are produced. The threshold value of $n_* /
n_e$ at 169 MHz is $n_*/n_e = 2.2 \times 10^{-7}$, corresponding to
$T_{b} \sim 10^{10}\,$K. Statistics of noise storm continuum brightness
temperatures are most abundant at 169 MHz, and Kerdraon \& Mercier
(1983) find that the brightest noise storm continua at that frequency
have $T_{b} \sim 5 \times 10^9\,$ -- $10^{10}\,$K. Accordingly, we consider
the value $n_*/n_e = 2.2 \times 10^{-7}$ in our calculations. However, some
authors (e.g., Klein 1995 and references therein) use a frequency-independent
value of $n_*/n_e = 10^{-5}$, and consequently we also consider this value
in our calculations (see Table~1).

It is convenient to treat the high-energy power law index, $\alpha_2$,
as a free parameter in our model. According to equation~(\ref{eq16}), in
order to obtain a finite value for the nonthermal electron energy
density $U_*$, we must have $\alpha_2 < -5$. Combining this constraint
with equation~(\ref{eq13}), we find that
\begin{equation}
D_0 \, \tau \ < \ {1 \over 10} \ .
\label{eq21b}
\end{equation}
The same condition can also be derived by noting that the second term in
equation~(\ref{eq16b}) for $U_*$ diverges in the limit $D_0 \, \tau \to
1/10$. Using equation~(\ref{eq9}) to substitute for $D_0$ in terms of
the stochastic acceleration timescale $t_{\rm accel}$ yields the
equivalent result
\begin{equation}
{\tau \over t_{\rm accel}} < {4 \over 5} \ .
\label{eq21c}
\end{equation}
Any steady-state physical configuration governed by the transport
equation~(\ref{eq6}) must satisfy this condition. Once the value of
$\alpha_2$ is selected, then $\alpha_1$ and the product $D_0 \tau$ can
be computed using equations~(\ref{eq13}). By combining this information
with observational estimates for the coronal temperature $T$ and the
ratio $n_* / n_e$, we can compute $p_c$ using equation~(\ref{eq15b}),
which is an implicit equation for $\xi_c = p_c^2/(2 m_e k T)$. We remind
the reader that $p_{c}$ signifies the critical momentum above which
stochastic acceleration dominates over losses on average (see
eq.~[\ref{eq11}]). In our example calculations we use $T =
10^{6}\,$K for the coronal temperature and either $n_*/n_e = 2.2
\times 10^{-7}$ or $n_*/n_e = 10^{-5}$ for the fraction of nonthermal
electrons.

Although noise storm continua are observed between $\sim 50$ -- $300$
MHz, observations are most abundant at 169 MHz (Kerdraon \& Mercier
1983). In what follows, we will be using observations at 169 MHz to
define the typical source size of noise storm continua. We therefore
utilize a value for the thermal electron density $n_{e}$ that
corresponds to a plasma frequency of 169 MHz (e.g., Krall \& Trivelpiece
1986),
\begin{equation}
n_e = 3.54 \times 10^{8} \, {\rm cm^{-3}}
\ .
\label{eq20}
\end{equation}
With $p_c$ already determined using equation~(\ref{eq15b}) along with
selected values for $n_*/n_e$ and $T$ as explained above, we can now
solve for the Fermi acceleration constant $D_0$ by using
equation~(\ref{eq11}) to write
\begin{equation}
D_0 = 4.41 \times 10^{-63} \, n_e \, p_c^{-3}
\ .
\label{eq21}
\end{equation}
Finally, the stochastic energization rate per unit volume is obtained
by combining equations~(\ref{eq16b}) and (\ref{eq17}).

The typical size of a noise storm continuum source at 169 MHz is $3^{'}
\sim 10^{10}\,$cm (Kerdraon \& Mercier 1983). The vertical extent of a
noise storm continuum source is not directly known, but the typical
relative bandwidth of the observed emission is $\delta f/f \sim
100\,\%$. If $H = n_e\,(\nabla\,n_e)^{-1}$ is the scale height of
density variation in the corona, the noise storm continuum emission must
emanate from a range of heights $(\delta f/f)\,H/2$ (e.g., Melrose
1980). Using $H \sim 10^5\,$km, this yields an estimate of
\begin{equation}
V \sim 10^{30} \ {\rm cm^{3}}
\label{eq22}
\end{equation}
for the volume of the acceleration region. This estimate assumes that
the noise storm emitting region is a cylinder of height $H/2$ and cross
section $3^{'} \times 3^{'}$, which is consistent with the data at 169
MHz. However, the cross section of the emitting region may vary with
frequency, and therefore with height in the corona, in which case
the emission volume would not be a simple cylinder. With this caveat in
mind, we shall adopt the simple cylindrical picture here and use a
value of $n_{e}$ referenced to 169 MHz because noise storm continua
statistics are most abundant at this frequency.

The power input to the nonthermal electron acceleration process is
computed using
\begin{equation}
L_{\rm in} = V\,dU_{*}/dt\,\,\,\,{\rm erg\,s^{-1}}\, , 
\label{eq23}
\end{equation}
where $dU_{*}/dt$ is given by equations~(\ref{eq16b}) and (\ref{eq17}).
Representative results are presented in Table~1.
\begin{table*}
\begin{tabular}{ccccccc}
	\hline
$n_{*}/n_{e}$ & $\alpha_{2}$ & $\epsilon_{c}$ & $D_{0}$ & $L_{\rm in}$ & $L_{\rm out}$ & $\eta$ \\ 
$2.2 \times 10^{-7}$ & -5.1 & 1.43 & 0.185 & $5.62 \times 10^{24}$ & $10^{17}$--\,$10^{18}$ & $\sim 10^{-8}$--\,$10^{-7}$ \\
$2.2 \times 10^{-7}$ & -6.0 & 1.42 & 0.186 & $8.00 \times 10^{23}$ & $10^{17}$--\,$10^{18}$ & $\sim 10^{-7}$--\,$10^{-6}$ \\
$1.0 \times 10^{-5}$ & -5.1 & 1.09 & 0.270 & $3.00 \times 10^{26}$ & $10^{17}$--\,$10^{18}$ & $\sim 10^{-10}$--\,$10^{-9}$ \\
$1.0 \times 10^{-5}$ & -6.0 & 1.08 & 0.280 & $4.2 \times 10^{25}$ & $10^{17}$--\,$10^{18}$ & $\sim 10^{-9}$--\,$10^{-8}$ \\
\hline
\end{tabular}
\caption[]{Representative results}
% The endnotes section will be placed here.
\theendnotes
The critical energy $\epsilon_{c}$ (eq.~[\ref{eq10}]) is expressed here in
keV. The quantity $D_{0}$ (eq.~[\ref{eq21}]) is in units of ${\rm
s}^{-1}$. The input power to the electrons $L_{\rm in}$ (eq.~[\ref{eq23}])
and the power observed in electromagnetic radiation $L_{\rm out}$ are
expressed in units of ${\rm erg\,s^{-1}}$.
\end{table*}
We find that $L_{\rm in} \sim 10^{23-26}$ ${\rm
erg\,s^{-1}}$. Expectedly, $L_{\rm in}$ is larger for the calculations where we have used the relatively larger
value of $n_{*}/n_{e}$. From independent observational considerations, the power
in noise storm continua is estimated to be $L_{\rm out} \sim 10^{17-18}
\,{\rm erg\,s^{-1}}$ (e.g., Elgaroy 1977; Raulin \& Klein 1994). The
associated efficiency,
\begin{equation}
\eta \equiv {L_{\rm out} \over L_{\rm in}} \ ,
\label{eq24}
\end{equation}
is therefore estimated to be in the range $10^{-10} \lsim \eta \lsim 10^{-6}$.

\section{Summary and Discussion}

We have stipulated a generic stochastic Fermi acceleration mechanism for
generating the nonthermal electrons responsible for noise storm
continua. The mathematical approach is based on a rigorous derivation of
the Green's function describing the acceleration of (initially)
monoenergetic electrons, which achieve a power-law distribution at high
energies (see eq.~[\ref{eq12}]). The Green's function is convolved with
the high-energy portion of the electron Maxwellian in the corona to
obtain the particular solution for the momentum distribution of the
nonthermal electrons responsible for producing the observed noise storm
emission (eq.~[\ref{particular}]). Integration of the particular solution
in turn yields exact solutions for the number and energy densities of
the nonthermal electrons, as well as an expression for the rate of
change of their energy density (eqs.~[\ref{eq15b}] -- [\ref{eq17}]).

Our work utilizes estimates for the ratio of the nonthermal to thermal
electron densities $n_*/n_e$ from Thejappa (1991) (implied by the
typically observed values for the noise storm continuum brightness
temperature $T_b$ at $\sim$ 169 MHz) and Klein (1995). Since we do not
have reliable observational estimates for the residence time $\tau$ of
the electrons in the acceleration region, we instead parametrize the
model using the high-energy power-law index $\alpha_{2}$ of the electron
distribution function. We must restrict ourselves to values of
$\alpha_{2} < -5$ in order to obtain a finite value for the nonthermal
electron energy density. By combining an observational value for
$n_*/n_e$ with a selected value for $\alpha_2$, we can use
equation~(\ref{eq15b}) to determine the critical momentum $p_c$, which
separates the thermal and nonthermal portions of the electron
distribution. The analytical solution for the energy density of the
nonthermal electrons is then used to compute the energization rate due
to second-order Fermi acceleration (see eqs.~[\ref{eq16b}] and
[\ref{eq17}]).

Typical observational values for the total electron number density $n_e$
and the volume of the noise storm emitting region $V$ were used to
arrive at the results presented in Table~1. These results demonstrate
that the power input to the electron acceleration process is around
$10^{23-26} \, {\rm erg\,s^{-1}}$, which is 6--10 orders of magnitude
larger than the power that is ultimately observed in the noise storm
continuum radiation. The efficiency of the process, starting from the
acceleration of the nonthermal electrons and culminating in the
observable noise storm continuum radiation, is therefore in the range
$10^{-10} \lsim \eta \lsim 10^{-6}$. Using data from type III decametric
storms and impulsive 2--10 keV electron events at 1~AU associated with
type I noise storms, Lin (1985) estimates that the energy release rate
for these electrons is around $10^{23}\,{\rm erg\,s^{-1}}$ (cf. also
Jackson \& Leblanc 1991). Klein (1995) estimates that the energy supply
to the energetic electrons is of the order of $10^{23-24} \, {\rm
erg\,s^{-1}}$. Prior to the calculations presented here, these were the
only (indirect, and rather approximate) estimates of the power input to
nonthermal electrons required to drive noise storm radiation.

We believe that the general analysis of the energy budget presented in
this paper, based on a generic second-order Fermi energization
mechanism, will help to guide subsequent work on the detailed
acceleration processes responsible for powering the noise storm
continua. Our results also provide an important quantitative data point
for discussions of electron acceleration in other high-temperature
astrophysical environments.

\acknowledgements
The authors are grateful to the anonymous referee who provided
a number of useful comments and suggestions which helped to
improve the manuscript. PS also acknowledges useful discussions
with Prof. Monique Pick.

\end{article}
\end{document}